\documentclass[a4paper,11pt]{article}
\usepackage[textwidth=15.2cm,textheight=22cm]{geometry}
\usepackage{amsmath,amsfonts,amssymb,color,psfrag,graphicx,fancyhdr,array}
\usepackage[latin1]{inputenc}

\usepackage{bbm}
\usepackage{amssymb}
\usepackage{amsmath}
\usepackage{amscd}
\usepackage{pst-all}
\usepackage{pst-node}

\usepackage{ifthen}

\def\a{\alpha}

\def\g{\gamma}

\def\e{\eta}
\def\eps{\epsilon}
\def\ve{\varepsilon}

\def\t{\tau}

\def\la{\lambda}

\newcommand{\ZC}{\mathbbm C}

\newcommand{\CE}{\mathcal{E}}       

\newcommand{\CL}{\mathcal{L}}       
\newcommand{\CW}{\mathcal{W}}
\newcommand{\CV}{\mathcal{V}}

\def\_{\;\;}
\def\^{\wedge}

\def\eqn#1{eq.~(\ref{#1})}

\def\eqns#1#2{eqs.~(\ref{#1}) and~(\ref{#2})}

\def\sfrac#1#2{{\textstyle\frac{#1}{#2}}}
\def\>{\rangle}
\def\<{\langle}
\def\+{\dagger}
\def\={\ =\ }
\def\tree{{\rm tree}}

\def\and{\qquad\textrm{and}\qquad}

\def\res#1{\ensuremath{\lbrace#1\rbrace}}
\def\ki#1{[\![#1]\!]}
\newcommand{\Nf}{\ensuremath{\mathcal N{=}4}\ }

\newcommand{\etal}{\emph{et al.\ }}
\newcommand{\entspricht}{\mathrel{\widehat{=}}}

\newcommand{\nnl}{\nonumber\\}
\newcommand{\half}{\sfrac{1}{2}}

\def\BoxOneMass#1#2#3#4#5{
\begin{pspicture}(30,30)
 \psframe[dimen=middle](10,10)(20,20)
 \SpecialCoor
  \psline(20,20)(23.53,23.53)
  \ifthenelse{\equal{#1}{}}{}{\uput*[45](20,20){\small #1}}
  \psline(20,10)(23.53,6.47)
  \ifthenelse{\equal{#2}{}}{}{\uput*[-45](20,10){\small #2}}
  \psline(10,10)(6.47,6.47)
  \ifthenelse{\equal{#3}{}}{}{\uput*[-135](10,10){\small #3}}
  \psline(10,20)(5.31,21.71)
  \ifthenelse{\equal{#4}{}}{}{\uput*[160](10,20){\small #4}}
 \psline(10,20)(8.29,24.69)
 \ifthenelse{\equal{#5}{}}{}{\uput*[110](10,20){\small #5}}
 \psdot(!10 3 60 cos mul sub 20 3 60 sin mul add)\psdot(!10 3 45 cos mul sub 20
3 45 sin mul add)\psdot(!10 3 30 cos mul sub 20 3 30 sin mul add)
 \NormalCoor
\end{pspicture}}

\def\BoxTwoMassEasy#1#2#3#4#5#6{
\begin{pspicture}(30,30)
 \psframe[dimen=middle](10,10)(20,20)
 \SpecialCoor
  \psline(20,20)(23.53,23.53)
  \ifthenelse{\equal{#1}{}}{}{\uput*[45](20,20){\small #1}}
 \psline(20,10)(24.69,8.29)
 \ifthenelse{\equal{#2}{}}{}{\uput*[-20](20,10){\small #2}}
  \psline(20,10)(21.71,5.31)
  \ifthenelse{\equal{#3}{}}{}{\uput*[-70](20,10){\small #3}}
  \psdot(!20 3 60 cos mul add 10 3 60 sin mul sub) \psdot(!20 3 45 cos mul add
10 3 45 sin mul sub) \psdot(!20 3 30 cos mul add 10 3 30 sin mul sub)
  \psline(10,10)(6.47,6.47)
  \ifthenelse{\equal{#4}{}}{}{\uput*[-135](10,10){\small #4}}
  \psline(10,20)(5.31,21.71)
  \ifthenelse{\equal{#5}{}}{}{\uput*[160](10,20){\small #5}}
 \psline(10,20)(8.29,24.69)
 \ifthenelse{\equal{#6}{}}{}{\uput*[110](10,20){\small #6}}
 \psdot(!10 3 60 cos mul sub 20 3 60 sin mul add)\psdot(!10 3 45 cos mul sub 20
3 45 sin mul add)\psdot(!10 3 30 cos mul sub 20 3 30 sin mul add)
 \NormalCoor
\end{pspicture}}

\def\BoxTwoMassHard#1#2#3#4#5#6{
\begin{pspicture}(30,30)
 \psframe[dimen=middle](10,10)(20,20)
 \SpecialCoor
  \psline(20,20)(23.53,23.53)
  \ifthenelse{\equal{#1}{}}{}{\uput*[45](20,20){\small #1}}
  \psline(20,10)(23.53,6.47)
  \ifthenelse{\equal{#2}{}}{}{\uput*[-45](20,10){\small #2}}
 \psline(10,10)(8.29,5.29)
 \ifthenelse{\equal{#3}{}}{}{\uput*[-110](10,10){\small #3}}
  \psline(10,10)(5.29,8.29)
  \ifthenelse{\equal{#4}{}}{}{\uput*[-160](10,10){\small #4}}
 \psdot(!10 3 60 cos mul sub 10 3 60 sin mul sub)
 \psdot(!10 3 45 cos mul sub 10 3 45 sin mul sub)
 \psdot(!10 3 30 cos mul sub 10 3 30 sin mul sub)
 \psline(10,20)(5.31,21.71)
 \ifthenelse{\equal{#5}{}}{}{\uput*[160](10,20){\small #5}}
  \psline(10,20)(8.29,24.69)
  \ifthenelse{\equal{#6}{}}{}{\uput*[110](10,20){\small #6}}
 \psdot(!10 3 60 cos mul sub 20 3 60 sin mul add)\psdot(!10 3 45 cos mul sub 20
3 45 sin mul add)\psdot(!10 3 30 cos mul sub 20 3 30 sin mul add)
  \NormalCoor
\end{pspicture}}

\def\BoxThreeMass#1#2#3#4#5#6#7{
\begin{pspicture}(30,30)
 \psframe[dimen=middle](10,10)(20,20)
 \SpecialCoor
  \psline(20,20)(23.53,23.53)
  \ifthenelse{\equal{#1}{}}{}{\uput*[45](20,20){\small #1}}
 \psline(20,10)(24.69,8.29)
 \ifthenelse{\equal{#2}{}}{}{\uput*[-20](20,10){\small #2}}
  \psline(20,10)(21.71,5.31)
  \ifthenelse{\equal{#3}{}}{}{\uput*[-70](20,10){\small #3}}
  \psdot(!20 3 60 cos mul add 10 3 60 sin mul sub) \psdot(!20 3 45 cos mul add
10 3 45 sin mul sub) \psdot(!20 3 30 cos mul add 10 3 30 sin mul sub)
 \psline(10,10)(8.29,5.29)
 \ifthenelse{\equal{#4}{}}{}{\uput*[-110](10,10){\small #4}}
  \psline(10,10)(5.29,8.29)
  \ifthenelse{\equal{#5}{}}{}{\uput*[-160](10,10){\small #5}}
 \psdot(!10 3 60 cos mul sub 10 3 60 sin mul sub)
 \psdot(!10 3 45 cos mul sub 10 3 45 sin mul sub)
 \psdot(!10 3 30 cos mul sub 10 3 30 sin mul sub)
 \psline(10,20)(5.31,21.71)
 \ifthenelse{\equal{#6}{}}{}{\uput*[160](10,20){\small #6}}
 \psline(10,20)(8.29,24.69)
 \ifthenelse{\equal{#7}{}}{}{\uput*[110](10,20){\small #7}}
 \psdot(!10 3 60 cos mul sub 20 3 60 sin mul add)\psdot(!10 3 45 cos mul sub 20
3 45 sin mul add)\psdot(!10 3 30 cos mul sub 20 3 30 sin mul add)
 \NormalCoor
\end{pspicture}}

\def\BoxFourMass#1#2#3#4#5#6#7#8{
\begin{pspicture}(30,30)
 \psframe[dimen=middle](10,10)(20,20)
 \SpecialCoor
 \psline(20,20)(21.71,24.69)
 \ifthenelse{\equal{#1}{}}{}{\uput*[70](20,20){\small #1}}
 \psline(20,20)(24.69,21.71)
 \ifthenelse{\equal{#2}{}}{}{\uput*[20](20,20){\small #2}}
 \psdot(!20 3 60 cos mul add 20 3 60 sin mul add) \psdot(!20 3 45 cos mul add 20
3 45 sin mul add) \psdot(!20 3 30 cos mul add 20 3 30 sin mul add)
 \psline(20,10)(24.69,8.29)
 \ifthenelse{\equal{#3}{}}{}{\uput*[-20](20,10){\small #3}}
  \psline(20,10)(21.71,5.31)
  \ifthenelse{\equal{#4}{}}{}{\uput*[-70](20,10){\small #4}}
  \psdot(!20 3 60 cos mul add 10 3 60 sin mul sub) \psdot(!20 3 45 cos mul add
10 3 45 sin mul sub) \psdot(!20 3 30 cos mul add 10 3 30 sin mul sub)
 \psline(10,10)(8.29,5.29)
 \ifthenelse{\equal{#5}{}}{}{\uput*[-110](10,10){\small #5}}
  \psline(10,10)(5.29,8.29)
  \ifthenelse{\equal{#6}{}}{}\uput*[-160](10,10){\small #6}
 \psdot(!10 3 60 cos mul sub 10 3 60 sin mul sub)
 \psdot(!10 3 45 cos mul sub 10 3 45 sin mul sub)
 \psdot(!10 3 30 cos mul sub 10 3 30 sin mul sub)
  \psline(10,20)(5.31,21.71)
  \ifthenelse{\equal{#7}{}}{}{\uput*[160](10,20){\small #7}}
  \psline(10,20)(8.29,24.69)
  \ifthenelse{\equal{#8}{}}{}{\uput*[110](10,20){\small #8}}
 \psdot(!10 3 60 cos mul sub 20 3 60 sin mul add)\psdot(!10 3 45 cos mul sub 20
3 45 sin mul add)\psdot(!10 3 30 cos mul sub 20 3 30 sin mul add)
 \NormalCoor
\end{pspicture}}

\numberwithin{equation}{section}

\begin{document}
\begin{titlepage}
\renewcommand{\thefootnote}{\fnsymbol{footnote}}
AEI-2010-077\hfill ITP-UH-05/10\vskip 2cm
\centerline{\Large{\bf {Dual conformal
constraints and infrared equations}}}
\centerline{\Large{\bf {from global residue theorems
in \Nf SYM theory}}} \vskip 1.0cm \centerline{Johannes Br\"{o}del${}^{a,b}$ and Song He${}^{a}$
}
\vskip .5cm
\centerline{${}^a$ \it {Max-Planck-Institut f\"{u}r
Gravitationsphysik}}
\centerline{\it {Albert-Einstein-Institut, Golm, Germany}}
\vskip .5cm
\centerline{${}^b$ \it {Institut f\"ur Theoretische Physik}}
\centerline{\it {Leibniz Universit\"at Hannover, Germany}}
\vskip 1.5cm
\centerline{\bf {Abstract}}
\vskip .5cm
Infrared equations and dual conformal constraints arise as consistency conditions on
loop amplitudes in \Nf super Yang-Mills theory. These conditions are linear relations between 
leading singularities, which can be computed in the Grassmannian formulation of \Nf super Yang-Mills 
theory proposed recently. Examples for infrared equations have been shown to be implied 
by global residue theorems in the Grassmannian picture.

Both dual conformal constraints and 
infrared equations are mapped explicitly to global residue theorems for one-loop 
next-to-maximally-helicity-violating amplitudes. In addition, the identity relating the BCFW and 
its parity-conjugated form of tree-level amplitudes, is shown to emerge from a particular
combination of global residue theorems. 
\vfill
email:\,\texttt{jbroedel@aei.mpg.de, songhe@aei.mpg.de}
\end{titlepage}


\section{Introduction}
Scattering amplitudes in maximally supersymmetric super
Yang-Mills (SYM) theory, possess many beautiful properties which are
obscured in their local formulations \cite{MangPark-Dixon-BDK}.
Remarkably simple formul\ae~for MHV amplitudes have been known since
\cite{Parke:1986gb,Nair:1988bq}. More recently, Witten's seminal
work on twistor string theory for \Nf SYM \cite{Witten:2003nn} has
triggered the development and exploration of many new techniques for
efficiently computing scattering amplitudes, such as MHV diagrams
\cite{CSW} and the Britto-Cachazo-Feng-Witten (BCFW) recursion relations \cite{BCF,BCFW}. In
addition, organizing amplitudes in a maximally supersymmetric way
has lead to the discovery of dual superconformal symmetry in \Nf SYM
theory \cite{Drummond:2007aua,Brandhuber:2007yx,Drummond:2008vq},
which can be combined with conventional superconformal symmetry to
yield the Yangian symmetry \cite{Plefka}. Recently, in
\cite{ArkaniHamed:2009dn, Mason:2009qx} the Grassmannian conjecture
was put forward: the Grassmannian integral was argued to encode many
of the previously hidden structures and symmetries of \Nf SYM
amplitudes.

The Grassmanian integral is conjectured to generate all-loop leading
singularities of \Nf SYM theory. Leading
singularities~\cite{BC,CK,Cachazo:2008vp,Cachazo:2008hp} are the
highest codimension singularities at $l$-loop level, obtained by
cutting $4l$ propagators in the generalized unitarity
method~\cite{BDK,quadruple}. The leading singularities\footnote{As
common in the literature we will use the terminology ``leading
singularity'' to refer to the discontinuity across a leading
singularity.} are well-defined and infrared-finite objects. It turns out 
that the box coefficients from expanding
a one-loop amplitude in \Nf SYM into scalar box functions are exactly
the one-loop leading singularities. Generalizing this fact, it was
conjectured that leading singularities together with their
corresponding integral basis can determine complete amplitudes of
\Nf SYM theory at any loop order \cite{ArkaniHamed:2008gz}.

There is evidence that leading singularities at all loop-orders are
related to each other by infrared (IR) equations
\cite{Kunszt:1994mc,Catani:1998bh,Catani:1999ss}. For \Nf SYM amplitudes at one-loop
level, IR equations can be derived by comparing the IR-divergences
of the amplitude with those of its expansion into box functions, as
will be reviewed in subsection \ref{sec:spaceform}. They
turn out to be simple linear relations among various box
coefficients (one-loop leading singularities) and the tree amplitude,
which can be viewed as special version of a leading singularity.

Furthermore, based on anomalous dual conformal symmetry, constraints
for one-loop leading singularities have been derived in
\cite{Brandhuber:2009xz,Elvang:2009ya}. While box coefficients as
well as tree amplitudes are covariant under dual conformal symmetry,
one-loop amplitudes exhibit a dual conformal anomaly due to infrared
divergences. Dual conformal constraints can be derived -- very
similarly to IR equations -- by comparing the anomaly of the complete
one-loop amplitude to the anomalies of the box functions. Not
surprisingly, these constraints, also being linear relations among
box coefficients and tree amplitudes, have been found to imply all
one-loop IR equations. In addition, there are new relations which
originate in the anomalies of the dual superconformal symmetry
exclusively.

In \cite{ArkaniHamed:2009dn} evidence has been put forward that
one-loop IR equations of \Nf SYM can be traced back to global residue
theorems (GRT) in the Grassmannian description of \Nf SYM theory. In this
formalism, leading singularities are expressed as residues of the
multi-dimensional complex contour integral~\cite{ArkaniHamed:2009dn}
\begin{equation}\label{eqn:Grass1}
\CL_{n;k}(\mathcal{W}_a) = \int \frac{d^{k \times n} C_{\a
a}}{(12\cdots k) \, (23\cdots (k+1) \,) \, \cdots (n 1 \cdots (k-1)
\,)} \prod_{\a = 1}^k \delta^{4|4}(C_{\alpha a} \mathcal{W}_a)\,.
\end{equation}
In analogy to Cauchy's theorem in the complex plane, where the
residues enclosed in a certain contour sum up to zero, there are
GRTs in the multidimensional complex space relating the residues of
\eqn{eqn:Grass1}. Starting from known IR equations and expressing
the leading singularities in terms of residues, it was shown for a
couple of examples that those equations indeed can be traced back to
GRTs. Despite of these promising results, a general map between IR
equations and GRTs has been missing so far. In this article, we will
propose that not only IR equations, but in fact all one-loop dual
conformal constraints, have their origin in GRTs in the Grassmannian
formulation.

Although we conjecture dual conformal constraints (and thus IR equations) to be related to GRTs
for any one-loop amplitude, we will limit our considerations to the NMHV sector.
In this sector any integration contour for the evaluation of the Grassmannian
integral is in one-to-one correspondence with a certain
choice of denominator factors in \eqn{eqn:Grass1} to be set to zero. Starting from the
N${}^2$MHV level, a choice of vanishing minors does not determine a
residue uniquely. Thus this identification can not be made
straightforwardly any more. In addition, in the NMHV situation a
definite map between one-loop leading singularities and residues is
known, while beyond NMHV a complete identification has not yet been
achieved.

It should be noted that only a subset of all available GRTs is used
to derive all one-loop dual conformal constraints. Since it is now
clear~\cite{Kaplan:2009mh,Bullimore:2009cb} that residues in
Grassmannian formulation correspond to all-loop leading
singularities, some GRTs should have interpretations as relations
involving higher-loop leading singularities. 
In order to map all one-loop constraints onto leading singularities, we
need to decide which residue occurs as a contribution to a 
leading singularity at a certain loop level \textit{first}. Here we will use 
the \textit{invariant label} introduced below as a criterion, while a similar
classification has been performed by considering the twistor support of leading
singularities in~\cite{Bullimore:2009cb}. In the same reference it was 
shown that NMHV leading singularities can maximally occur at three-loop level.

By systematically translating all box coefficients appearing in
one-loop dual conformal constraints into
residues in the Grassmannian formulation, we find a general map between
all one-loop dual conformal constraints and combinations of GRTs in
the NMHV sector. Additionally, we also propose a general formula to
assign a sum of GRTs to any one-loop IR equation. Finally, the mapping is rounded off
by identifying a mechanism which provides the highly nontrivial identities
relating the BCFW and the P(BCFW) representations of the NMHV tree
amplitude. These identities ensure the absence of
spurious poles, parity invariance and cyclicity of tree amplitudes~\cite{ArkaniHamed:2009dn}.

It is difficult to generalize the result to higher-loop level. The
absence of a general notion for an integral basis, already seen at two loops,
results in the lack of a clear identification of dual conformal
constraints and IR equations beyond one-loop. While there are
definitely additional relations originating in GRTs which correspond
to these higher-loop constraints, the identification and exploration
of those structures is left for a future project.

After describing the framework and introducing the conventions in
section~\ref{sec:basics}, we will proceed to classify residues by the invariant label in
section~\ref{sec:classification}. The mapping of one-loop IR
equations and -- more generally -- dual conformal constraints to GRTs
will be presented in section~\ref{sec:mapping}. 


\section{Prerequisites}\label{sec:basics}


\subsection{Spacetime formulation}\label{sec:spaceform}
We are going to use the \Nf on-shell super space introduced in~\cite{Nair:1988bq} 
with Grassmann coordinates
$\e$ and the usual spinor-helicity variables $\la,\tilde{\la}$.
Kinematical invariants are defined as
\begin{equation}\label{eqn:kininv}
\ki{i}_m=\,(k_i+k_{i+1}+\cdots+k_{i+m-1})^2\,,
\end{equation}
with all momenta being on-shell. This results in
$\ki{1}_2=s_{12}=2 k_1\cdot k_2$ and $[\![2]\!]_3=t_{234}=2(
k_2\cdot k_3+ k_2\cdot k_4+ k_3\cdot k_4)$ for example. All
considerations below are given in a completely super-symmetric
invariant form. However, since 4-mass box integrals are IR finite
and exhibit no conformal anomaly, they never participate in one-loop IR
equations or dual conformal constraints.

As is well known, one-loop superamplitudes in \Nf SYM theory can be
expanded into scalar box integrals. However, both in the context of
leading singularities and dual conformal symmetry, it is most
natural and convenient to use box functions as the basis,
\begin{equation}\label{eqn:expansion}
 M^{\rm 1\hbox{-}loop}=\sum_i C_iF_i,
\end{equation}
where $F_i=-\sfrac{I_i}{2\sqrt{R_i}}$ are IR-divergent box
functions, simply related to scalar box integrals $I_i$ by kinematic
factors $R_i$ (see the first paper of ref.~\cite{BDK} for details). Box functions can be identified with certain box
configurations. 
The quantities $C_i$ are called box coefficients, which are exactly one-loop leading 
singularities.

Considering whether there are one or more legs attached to the corners of a
box diagram naturally leads to the following categories:
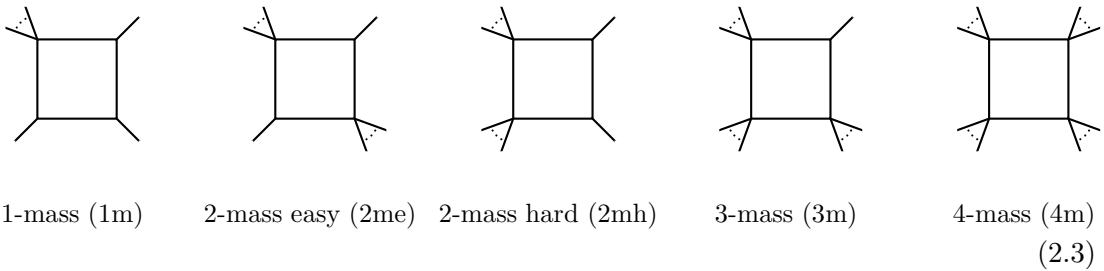
\begin{equation}
\psset{xunit=3pt,yunit=3pt,linewidth=0.8pt,labelsep=16pt,dotsize=0.8pt}
\begin{pspicture}(150,33)
\rput[b](15,3){\BoxOneMass{\ }{\ }{\ }{\ }{\ }}
\rput[B](15,0){\small 1-mass (1m)}
\rput[b](45,3){\BoxTwoMassEasy{\ }{\ }{\ }{\ }{\ }{\ }}
\rput[B](45,0){\small 2-mass easy (2me)}
\rput[b](75,3){\BoxTwoMassHard{\ }{\ }{\ }{\ }{\ }{\ }}
\rput[B](75,0){\small 2-mass hard (2mh)}
\rput[b](105,3){\BoxThreeMass{\ }{\ }{\ }{\ }{\ }{\ }{\ }}
\rput[B](105,0){\small 3-mass (3m)}
\rput[b](135,3){\BoxFourMass{\ }{\ }{\ }{\ }{\ }{\ }{\ }{\ }}
\rput[B](135,0){\small 4-mass (4m)}
\end{pspicture}
\end{equation}

The corresponding box functions and their coefficients are
conveniently labeled by the distribution of legs onto the corners of
the boxes. Given the known number of legs $n$, it is sufficient to
note the first legs attached to the four corners of the box. If not
stated differently, the first entry of the four-element list
$r,s,t,u\in\{1,\ldots,n\}$ is assumed to label the massless leg in the
upper right corner. In some cases, despite being redundant, the type
of box will be noted as a superscript on the coefficient. Employing
those conventions, \eqn{eqn:expansion} reads
\begin{equation}\label{eqn:explicitexpansion}
M^{\rm 1\hbox{-}
loop}_n(\la,\tilde\la,\e,\eps)=\sum_{\{rstu\}}C_{rstu}(\la,\tilde\la,\e)F_{rstu}(\la,\tilde\la,\eps)\,,
\end{equation}
where the parameter of dimensional regularization is defined by $4-2\eps=D$.

In order to obtain the IR equations, one starts from the
IR-divergent part of dimensionally regularized one-loop amplitudes of
\Nf SYM~\cite{Kunszt:1994mc},
\begin{equation}\label{eqn:IR}
\left. M_n^{\rm 1 \hbox{-} loop} \right|_{IR} =
-\frac{r_\Gamma}{\eps^2}\sum_{i=1}^n (-\ki{i}_2)^{-\eps} M^{\tree},
\end{equation}
where
$r_\Gamma:=\Gamma(1+\epsilon)\Gamma^2(1-\epsilon)\Gamma(1-2\epsilon)$.

The IR-divergent part of box functions is given by \cite{Bern:1995ix}
\begin{eqnarray}\label{eqn:IRdivBoxes}
\left. F^{\rm 1m}(p,q,r,P) \right|_{IR} & = &
-\frac{r_\Gamma}{\eps^2}\left( (-s)^{-\eps} + (-t)^{-\eps} -
(-P^2)^{-\eps} \right) \nnl \left. F^{\rm 2me}(p,P,q,Q)
\right|_{IR}& = & -\frac{r_\Gamma}{\eps^2}\left( (-s)^{-\eps} +
(-t)^{-\eps} - (-P^2)^{-\eps} - (-Q^2)^{-\eps}\right) \nnl \left.
F^{\rm 2mh}(p,q,P,Q) \right|_{IR} & = &
-\frac{r_\Gamma}{\eps^2}\left(
\frac{1}{2}(-s)^{-\eps}+(-t)^{-\eps}-\frac{1}{2}(-P^2)^{-\eps}-\frac{1}{2}
(-Q^2)^{-\eps}\right)\nnl \left. F^{\rm 3m}(p,P,R,Q) \right|_{IR} &
= & -\frac{r_\Gamma}{\eps^2}\left( \frac{1}{2}(-s)^{-\eps} +
\frac{1}{2}(-t)^{-\eps}- \frac{1}{2}(-P^2)^{-\eps} -
\frac{1}{2}(-Q^2)^{-\eps}\right)\,,
\end{eqnarray}
where $F^{\rm 4m}$ has not been listed because of its IR finiteness.
In the notation $F(K_1,K_2,K_3,K_4)$ of \eqn{eqn:IRdivBoxes},
capital letters define sums of consecutive massless momenta and
lower case letters correspond to single (null) momenta. In addition,
the two main kinematical invariants are defined as $s=(K_1+K_2)^2$
and $t=(K_2+K_3)^2$.

IR equations are obtained by requiring the consistency between
\eqn{eqn:IR} and \eqn{eqn:IRdivBoxes}. There is one IR equation
for each kinematical invariant, whose IR behavior shall be considered. Thus each IR equation is
conveniently labeled by this kinematical invariant.
Since $\ki{i}_m=\ki{i+m}_{n-m}$ by momentum conservation, we can
limit the considerations to $2\leq m\leq \lfloor n/2\rfloor$.

After expanding the left hand side of \eqn{eqn:IR} using
\eqns{eqn:expansion}{eqn:IRdivBoxes}, there are two different
situations:
\begin{itemize}
\item if the considered kinematical invariant is of the form $\ki{i}_2$, the sum of
box-coefficients from the left hand side has to be proportional to
the tree amplitude $M^{\rm tree}$.
\item For kinematical invariants of the form $\ki{i}_m$ with $m>2$ there is no
contribution from the right-hand side of~\eqn{eqn:IR}. Thus the
total sum of box-coefficients will vanish, which leads to a relation
purely between box coefficients themselves.
\end{itemize}
The resulting IR equations are relations between IR-finite
quantities: the tree amplitude and one-loop leading singularities.
It is not difficult to count the number of IR equations for a
certain number of legs. There are $\sfrac{n(n-3)}{2}$ IR
equations in total, which split up into $n$ equations involving the
tree amplitude and $\sfrac{n(n-5)}{2}$ pure one-loop equations.

The IR equations considered above represent only a subset of all known
relations between one-loop leading singularities. The anomalous dual conformal symmetry at
one-loop level results in constraints that imply, but are not limited to,
the IR equations. Here we only state the result without repeating the
derivation, which can be found in~\cite{Brandhuber:2009xz,Elvang:2009ya}.

The dual conformal anomaly of one-loop amplitudes can be obtained by
applying the shifted dual conformal generator\footnote{Box coefficients 
are invariant~\cite{Plefka} under the shifted operator $\hat{K}^\mu$ rather than covariant 
under the usual $K^\mu$.} $\hat{K}^{\mu}$ to both sides
of~\eqn{eqn:expansion}. The dual conformal anomaly of one-loop
amplitudes has been conjectured in~\cite{Drummond:2008vq} to be
proportional to the tree amplitude. On the right-hand side, the
generator can pass through the box coefficients due to their
invariance. Thus $\hat{K}^{\mu}$ directly acts on the box
functions $F_{rstu}$, whose anomaly structure resembles that of
their infrared divergences. Finally one finds $n(n-4)$ independent
equations, \eqns{eqn:conformaltree}{eqn:E}, which fall into two categories.
\begin{itemize}
\item $n$ combinations of (2-mass hard and 1-mass) box coefficients
equal to tree amplitudes, which are in one-to-one correspondence with
the $n$ IR equations of this category,
\begin{equation}\label{eqn:conformaltree}
\CE_{i,i-2}=-\CE_{i-1,i}=-2M^{\rm tree}_n\, ,
\end{equation}
where $i=1,...,n$ and
\begin{equation}\label{eqn:conformalvanish}
\CE_{i,i-2}=-\CE_{i-1,i}:=-\sum_{j=i+1}^{i+n-3}C_{i-2,i-2,i,j}
\end{equation}
are boundary cases of the more general $\CE_{i,k}$ defined in \eqn{eqn:E} below.

\item $n(n-3)$ combinations of box coefficients vanish,
\begin{equation}
\CE_{i,k}=0\,,
\end{equation}
where $i=1,...,n$ and $k=i+2,...,i+n-3$. The quantities $\CE_{i,k}$ are
defined as
\begin{equation}\label{eqn:E}
\CE_{i,k}:=\sum_{j=k+1}^{i+n-2}C_{i,k,j,i-1}-\sum_{j=i+1}^{k-1}C_{i,j,k,i-1}\,.
\end{equation}
Since there are $2n$ algebraic identities among them, we have
$n(n-5)$ independent constraints. These constraints imply the
$\sfrac{n(n-5)}{2}$ IR equations of the same category but also
$\sfrac{n(n-5)}{2}$ new constraints.
\end{itemize}
Finally we would like to emphasize that, although in the following we
mainly focus on the NMHV sector, the IR equations and dual conformal
constraints are valid for any N${}^{k-2}$MHV amplitude.

\subsection{Grassmannian formulation}\label{sec:Grassform}
The object which was suggested by Arkani-Hamed \etal to calculate
leading singularities of all-loop $n$-particle amplitudes in the
N${}^{(k-2)}$MHV sector is a gauged version~\cite{ArkaniHamed:2009dn} of
\eqn{eqn:Grass1}
\begin{equation}\label{eqn:fGrassmannian}
{\CL}_{n;k} = L_{n;k} \times \delta^4(\sum_a p_a),
\end{equation}
where $L_{n;k}$ is defined as
\begin{equation}\label{eqn:fGrassexpl}
 L_{n;k} = J(\la,\tilde \la) \int
 \frac{d^{(k-2)\times(n-k-2)} \tau}{\left[(12\cdots k) \, (23\cdots
 (k+1)\, ) \, \cdots (n 1 \cdots (k-1)\,)\right](\tau)}
 \prod_I \delta^4(\tilde \e_I + c_{Ii}(\t) \tilde \e_i)\,.
\end{equation}
In the above expression, $J$ is a Jacobian prefactor depending on
the precise fixing of the gauge. The remaining variables to
integrate over are $\t_\g$, where $\g=1,\ldots,d$ with
$d:=(k-2)\times(n-k-2)$. The objects in the denominator, called
minors,
\begin{equation}\label{eqn:minors}
(m_1  \cdots m_k) \equiv \ve^{\a_1 \cdots \a_k}
C_{\a_1 m_1} \cdots C_{\a_k m_k}\,,
\end{equation}
are obtained from a gauge-fixed $k\times n$ matrix $C$ whose nontrivial elements
$c_{Ii}$ are solutions to the kinematical constraints
\begin{equation}
 \la_i-c_{Ii}(\t_\g)\la=0,\quad \tilde\la_I+c_{Ii}(\t_\g)\tilde\la_i=0.
\end{equation}

A couple of simplifications take place in the NMHV sector: the number of 
integration variables is just
$d=n-5$, and no composite residues occur~\cite{ArkaniHamed:2009dn}.
More importantly, the degree of the minors \eqn{eqn:minors} can be
shown to be $\min[k-2, n-k-2]$, which renders the NMHV situation the
easiest nontrivial one: for $k=3$ the minors are linear expressions
in $\t$. Note, that the integrand is a holomorphic function of the
(complexified) variables $\t_\g$. In this light,
\eqn{eqn:fGrassmannian} is a contour integral in $\ZC^{d}$.

A residue of \eqn{eqn:fGrassexpl} occurs at a multiple zero of
degree $d$. In the NMHV situation this is equivalent to choosing $d=n-5$ minors
to vanish simultaneously. Due to the linear degree of the minors there is 
exactly one solution and thus one residue for each of these choices. 
Correspondingly, any residue in the NMHV sector can be identified by noting
the vanishing minors, where each minor (cf. \eqn{eqn:minors}) is referred to by its
first entry $m_1$. The resulting list with $n-5$ elements is enclosed in curly brackets.

In the 8-point NMHV situation a residue can be calculated by
choosing $3=n-5$ minors to vanish. For example the residue obtained by setting the 
denominator terms $(123),\,(234)$ and $(781)$ to zero, will be referred 
to as $\lbrace1,2,7\rbrace$.

As shown in \cite{ArkaniHamed:2009dn}, the determination of a
multi-dimensional residue includes a determinant, which renders the
labeling for residues totally antisymmetric, for example
\begin{equation}
 \lbrace i,j,k\rbrace=-\lbrace i,k,j\rbrace\,.
\end{equation}

Although the labeling of NMHV residues with $n-5$ coordinates is
favorable for lower-point amplitudes, for general considerations we
will fall back to the \textit{complementary labeling}. The usual labeling can
be obtained from the 5-number complementary labeling by the bar
operation
\begin{equation}
\res{j_1,\ldots,j_{(n-5)}}=\overline{\res{i_1,\ldots,i_5}}=\res{\Xi}\cdot\text{sgn}(i_1,\ldots,i_5)\cdot\text{sgn}(i_1,\ldots,i_5,\Xi)
\end{equation}
where $\Xi$ is the ordered complement
\begin{equation}
\{1,\ldots,n\}\backslash\{i_1,\ldots,i_5\}\,.
\end{equation}

In the NMHV sector, there is a clear map between box
coefficients and residues. The simplest situation occurs for the
3-mass box, where the following box coefficient is given by:
\begin{equation}\label{eqn:3MBox}
\psset{xunit=3.6pt,yunit=3.6pt,linewidth=1.2pt,labelsep=21pt,dotsize=1.8pt}
\begin{pspicture}(30,33)
\rput[b](15,3){\BoxThreeMass{$1$}{$2$}{$s-1$}{$s$}{$t-1$}{$t$}{$n$}}
\rput[B](15,0){$C^{3m}_{12st}\entspricht\overline{\res{s-2,s-1,t-2,t-1,n}},$}
\end{pspicture}
\end{equation}
and any other 3-mass boxes can be obtained by cyclic shifts.

The expressions for other box coefficients can be easily obtained as
sums of (degenerate) 3-mass boxes by employing the results from
\cite{Drummond:2008bq}:
\begin{align}\label{eqn:otherBoxes}
C^{1m}_{r,r+1,r+2,r+3}\=&C^{2me}_{r+2,r+3,r,r+1}+C^{3m}_{r+1,r+2,r+3,r}\nnl
C^{2mh}_{r,r+1,r+2,s}\=&C^{3m}_{r+1,r+2,s,r}+C^{3m}_{r,r+1,r+2,s}
&(s>r+3,\,r>s+1)\nnl
C^{2me}_{r,r+1,s,s+1}\=&\sum_{\substack{u,v\\u\geq
r+2\\u+2\leq v\leq s}}C^{3m}_{r,r+1,u,v}+\sum_{\substack{u,v\\u\geq
s+2\\u+2\leq v\leq r}}C^{3m}_{s,s+1,u,v}&(s>r+2,r>s+2),
\end{align}
where all indices have to be understood modulo $n$. By writing `$>$' we mean
`$> \textrm{mod}\,n$' and the summations have also to be adapted
accordingly. If not stated otherwise, the modulo-$n$ notation will be understood implicitly below.

Finally, the NMHV tree-amplitude can be expressed in terms of
residues. With the sums
\begin{align}
 E&\=\sum_{k\text{ even}}\res{k}\nnl
 O&\=\sum_{k\text{ odd}}\res{k}
\end{align}
and the product\footnote{Here we use the notation: $\{i\}\{j\}:=\res{i,j}$ and $\{i+j\}\{k\}:=\res{i,j}+\res{i,k}$.}
\begin{equation}\label{eqn:starproduct}
 \res{i_1}\star\res{i_2}=\left\lbrace\begin{array}{l l}\res{i_1, i_2}\text{ if }\,\,i_1<i_2\\ 0\quad\text{otherwise}\end{array}\right\rbrace
\end{equation}
the BCFW form of the tree-amplitude is given by
\begin{equation}\label{eqn:BCFW}
 M^\tree_{\rm BCFW}=E\,\star\,O\,\star\,E\,\star\,\cdots\,,
\end{equation}
and the parity-conjugated (P(BCFW)) form is obtained from
\begin{equation}\label{eqn:PBCFW}
 M^\tree_{\rm P(BCFW)}=(-1)^{n-5}\,O\,\star\,E\,\star\,O\,\star\,\cdots\,.
\end{equation}

While for tree-level amlitudes and one-loop leading singularities there is a clear map to
residues, there is also evidence that certain residues encode higher-loop
information \cite{ArkaniHamed:2009dn,Kaplan:2009mh,Bullimore:2009cb}.
An interesting question is the following: is it possible to distinguish, which residues appear at
tree and one-loop level and which participate in higher-loop leading singularities only?
Section~\ref{sec:classification} is devoted to this classification.

As discussed in reference~\cite{ArkaniHamed:2009dn}, residues of
\eqn{eqn:fGrassexpl} are not independent objects, but are subject to
global residue theorems (GRTs). In particular, all NMHV GRTs can be
generated from basic GRTs, which have the form,
\begin{equation}
\sum_{j=1}^n\{j,i_1,...,i_{n-6}\}=0\,.
\end{equation}
where $(i_1,...,i_{n-6})$ is referred to as \textit{source term} and will be
used to uniquely label basic GRTs below. It is not difficult to see that any GRT
constrains the sum of six residues to vanish. In order to avoid confusion, source
terms are enclosed by usual brackets $()$, while residues will be enclosed
by curly brackets $\lbrace\rbrace$.

In reference \cite{ArkaniHamed:2009dn} it was shown on a couple of
examples that indeed one-loop IR equations and the identity between
the BCFW and P(BCFW) form of the tree amplitude can be traced back
to GRTs in the Grassmanian formulation. Since at the NMHV level one
can unambiguously map all one-loop leading singularities to
residues, it is natural to investigate the precise map between all
one-loop IR equations, and more generally, dual conformal constraints to sums of GRTs.
In addition, we will find out which GRTs imply the equality of the BCFW and the P(BCFW) form 
of the tree amplitude, which will be referred to as \textit{remarkable identity}. 
This analysis will be performed in section~\ref{sec:mapping}.



\section{Classification of residues}\label{sec:classification}

As we have discussed above, any residue at NMHV level is labeled by
$n-5$ numbers determining the minors to be set to zero in
\eqn{eqn:fGrassexpl}.

Employing the identifications \eqns{eqn:3MBox}{eqn:otherBoxes}, one
can single out all residues which appear in the one-loop leading
singularities. This will also include the constituents of the tree
amplitude, as those are related to the one-loop leading
singularities by \eqn{eqn:IR}. For $n\leq 7$ this covers all
possible residues.

Starting from $n=8$, there are residues
which only contribute to at least two-loop leading singularities, but
do not participate in any one-loop leading singularities \cite{ArkaniHamed:2009dn}.
However, residues occurring already in one-loop singularities can contribute
to two-loop (and higher) leading singularities.

As argued in \cite{Bullimore:2009cb}, certain residues appear at
three-loop level only for amplitudes with $n\geq 10$. This
is also the maximal loop-level for NMHV leading singularities: as
shown in the same reference, there are no leading singularities
at four-loops and higher. 

Assuming the Grassmannian conjecture to be true, it is clear that
residues for any NMHV amplitude should be classified by whether they
appear at the one-loop, two-loop or three-loop level \textit{first}. Although they can
contribute to higher-loop leading singularities, we will refer to those residues in a slightly
inaccurate manner as one-loop, two-loop and three-loop residues respectively.

As will be proven below, residues can be classified by using invariant labels. Given 
any sequence of cyclically ordered numbers $(i_1,\ldots,i_p)$
where $i_l\in\{1,\ldots,n\}$ and $p\leq n$, the invariant label is defined as the
set
\begin{equation}
 \lbrace(i_2-i_1)\,\text{mod}\,n,\,(i_2-i_1)\,\text{mod}\,n,\,\ldots,(i_1-i_p)\,
\text{mod}\,n\rbrace\,.
\end{equation}
As is obvious from the above definition, the name invariant label
refers to its invariance under cyclic shifts of $(i_1,\ldots,i_p)$.
Since any residue is a sequence of $n-5$ numbers from ${1,\ldots,n}$, one can
determine an invariant label for each of them.

Starting from the fact that any invariant label for a residue is a
decomposition of $n$ into $n-5$ numbers, it is easy to prove that,
given a sufficiently large $n$, there are exactly seven distinct
invariant labels:
\begin{center}
\begin{tabular}{ll}
\hline
type&invariant label\\
\hline
1&$\lbrace1,...,1,6\rbrace$\\
2&$\lbrace1,...,1,2,5\rbrace$\\
3&$\lbrace1,...,1,3,4\rbrace$\\
4&$\lbrace1,...,1,2,3,3\rbrace$\\
5&$\lbrace1,...,1,2,2,4\rbrace$\\
6&$\lbrace1,...,1,2,2,2,3\rbrace$\\
7&$\lbrace1,...,1,2,2,2,2,2\rbrace.$\\
\end{tabular}
\label{tab:decomposition}
\end{center}

In addition to the classification of the loop level, we will find the invariant
label to contain additional information: certain types of box coefficients
correspond to particular types of one-loop residues, as will be discussed below.

The mapping of box coefficients to residues is given in terms of the
complementary labeling defined in section~\ref{sec:Grassform}. Therefore, in a first step we will have to find classes
of complementary labels corresponding to types of residues.
As expected from the definition of the
invariant label, the criterion for classification is the number of succesive
subsequences. Again there are exactly seven types,
corresponding to the seven possible invariant labels in the table above:
\begin{center}
\begin{tabular}{lll}
\hline
type&complementary sequence& length of succ. subseq.\\[2pt]\hline\\[-12pt]
1&$\overline{\res{i,i+1,i+2,i+3,i+4}}$&5\\[2pt]\hline\\[-12pt]
2&\begin{minipage}{48mm}$\overline{\res{i,i+1,i+2,i+3,j_{>i+4}}}$\\
$\overline{\res{j,i_{>j+1},i+1,i+2,i+3}}$\end{minipage}&4\\[12pt]\hline\\[-12pt]
3&\begin{minipage}{48mm}$\overline{\res{i,i+1,i+2,j_{>i+3},j+1}}$\\
$\overline{\res{j,j+1,i_{>j+2},i+1,i+2}}$\end{minipage}&3 and 2\\[12pt]\hline\\[-12pt]
4&\begin{minipage}{48mm}$\overline{\res{i,i+1,j_{>i+2},j+1,k_{>j+2}}}$\\
$\overline{\res{k,i_{>k+1},i+1,j_{>i+2},j+1}}$\\$\overline{\res{i,i+1,k_{>i+2},j_{>k+1},j+1}}$\end{minipage}&2 and 2\\[20pt]\hline\\[-12pt]
5&\begin{minipage}{48mm}$\overline{\res{i,i+1,i+2,j_{>i+3},k_{>j+1}}}$\\
$\overline{\res{j,k_{>j+1},i_{>k+1},i+1,i+2}}$\\$\overline{\res{j,i_{>j+1},i+1,i+2,k_{>i+3}}}$\end{minipage}&3\\[20pt]\hline\\[-12pt]
6&\begin{minipage}{48mm}$\overline{\res{i,i+1,j_{>i+2},k_{>j+1},l_{>k+1}}}$\\
$\overline{\res{j,k_{>j+1},l_{>k+1},i_{>l+1},i+1}}$\end{minipage}&2\\[12pt]\hline\\[-12pt]
7&$\overline{\res{i,j_{>i+1},k_{>j+1},l_{>k+1},m_{>l+1}}}$&0
\end{tabular}
\end{center}

Comparing this table with the results for one-loop leading
singularities, \eqns{eqn:3MBox}{eqn:otherBoxes}, it is
straightforward to see that types 1 to 4 correspond to one-loop residues. The invariant labels
of types 1 to 4 contain exactly one even number.

Alternatively, we could have considered the BCFW and P(BCFW) form
of NMHV tree amplitudes given by \eqns{eqn:BCFW}{eqn:PBCFW}. Using
the fact that their residues are odd/even alternating sequences, it
follows immediately that their invariant labels can contain one even
number only. Since those representations of the tree amplitudes are
built from one-loop residues, we again arrive at the conclusion that
they belong to types 1 to 4.

Remembering the results from refs.~\cite{ArkaniHamed:2009dn,Bullimore:2009cb} stated at 
the beginning of this section, the remaining types $5$, $6$ and $7$ must correspond to higher-loop residues.
The fact that all leading singularities for $n\leq
7$ are combinations of one-loop residues nicely agrees with the lack
of decompositions of types 5, 6 or 7 at $n\leq 7$.

In addition, since types 5 and 6 can appear for $n\geq 8$ but the last
type only appears for $n\geq 10$, we can fit this fact to the
results of~\cite{Bullimore:2009cb}. We conjecture that types 5 and 6
correspond to two-loop residues while three-loop residues can be assigned
to type 7.

\begin{center}
\begin{small}
\begin{tabular}{ll}
\hline
$n$&invariant label\\
\hline
6&\res{6}\\
7&\res{1,6},\res{2,5},\res{3,4}\\
8&\res{1,1,6},\res{1,2,5},\res{1,3,4},\res{2,3,3},\res{2,2,4}\\
9&\res{1,1,1,6},\res{1,1,2,5},\res{1,1,3,4},\res{1,2,2,3},\res{1,2,2,4},\res{2,2,2,3}\\
10&\res{1,1,1,1,6},\res{1,1,1,2,5},\res{1,1,1,3,4},\res{1,1,2,2,3},\res{1,1,2,2,4},\res{1,2,2,2,3},\res{2,2,2,2,2}\\
\vdots&\vdots\\
n&\begin{minipage}{12cm}\res{1,...,1,6},\res{1,...,1,2,5},\res{1,...,1,3,4},\res{1,...,1,2,3,3},\\
                                       \res{1,...,1,2,2,4},\res{1,...,1,2,2,2,3},\res{1,...,1,2,2,2,2,2}\end{minipage}.
\end{tabular}\label{tab:decomposition2}
\end{small}
\end{center}

It is remarkable to see that no further types appear as the number
of particles increases, which agrees with the claim that all NMHV
leading singularities are combinations of these three types of
residues \cite{Bullimore:2009cb}.

If we examine the first four cases more carefully, we can see
further classifications among these one-loop residues. By comparing them
with results in section \ref{sec:Grassform}, we conclude that 3-mass
leading singularities can receive contributions from type 2 and 4
residues and residues for 2-mass hard leading singularities can be
type 1 and 3. For 2-mass easy, the corresponding residues are of type 2,
3 and 4, and 1-mass leading singularities can have all four possible
cases for one-loop residues. In summary, a complete classification of
residues based on invariant labels is presented in table below

\begin{center}
\psset{xunit=1.6pt,yunit=1.6pt,linewidth=0.6pt,labelsep=13pt,dotsize=1.0pt}
\begin{tabular}{ccccccc}
\hline
&3m&2mh&2me&1m&2-loop&3-loop\\[2pt]\hline\\[-12pt]
&\begin{pspicture}(17,10)
\rput[b](10,-14){\BoxThreeMass{}{}{}{}{}{}{}}
\end{pspicture}&
\begin{pspicture}(17,10)
\rput[b](10,-14){\BoxTwoMassHard{}{}{}{}{}{}}
\end{pspicture}&
\begin{pspicture}(17,10)
\rput[b](10,-14){\BoxTwoMassEasy{}{}{}{}{}{}}
\end{pspicture}&
\begin{pspicture}(17,10)
\rput[b](10,-14){\BoxOneMass{}{}{}{}{}}
\end{pspicture}&
\begin{pspicture}(22,10)
\rput[b](2,-6.3){
\psset{xunit=5pt,yunit=5pt,linewidth=0.6pt}
\psline{-}(0,5)(1.05,3.95)\psline{-}(0,0)(1.05,1.05)\psline{-}(6,0)(4.95,1.05)\psline{-}(6,5)(4.95,3.95)
  \psframe[dimen=middle](1,1)(3,4)\psframe[dimen=middle](3,1)(5,4)
}
\end{pspicture}&
\begin{pspicture}(30,10)
\rput[b](2,-6.3){
\psset{xunit=5pt,yunit=5pt,linewidth=0.6pt}
\psline{-}(0,5)(1.05,3.95)\psline{-}(0,0)(1.05,1.05)\psline{-}(8,0)(6.95,1.05)\psline{-}(8,5)(6.95,3.95)
  \psframe[dimen=middle](1,1)(3,4)\psframe[dimen=middle](3,1)(5,4)\psframe[dimen=middle](5,1)(7,4)
}
\end{pspicture}\\[12pt]\hline\\[-12pt]
types&
\begin{minipage}{17mm}\center\small{2 or 4}\end{minipage}&
\begin{minipage}{17mm}\center \small (1 + 3) or \\(3 + 3)\end{minipage} & 
\begin{minipage}{17mm}\center \small (1 + 2 +3)\\or (1 + 1)\end{minipage} & 
\begin{minipage}{17mm}\center\small{1,2,3,4}\end{minipage}&
\begin{minipage}{17mm}\center\small{5,6}\end{minipage}&
\begin{minipage}{17mm}\center\small{7}\end{minipage}
\end{tabular}.
\label{tab:types}
\end{center}


\section{Mapping one-loop dual conformal constraints and IR equations to GRTs}\label{sec:mapping}

After having classified all NMHV residues, all one-loop dual
conformal constraints will be translated into residues in this
section. First of all, the source terms of the GRTs necessary to
show
\begin{equation}\label{eqn:conformalnotree2}
\CE_{i,k}=0\,
\end{equation}
for $i=1,...,n$ and $k=i+2,...,i+n-3$ shall be investigated.

Starting with $\CE(1,4)=0$ for an amplitude with $n=9$ legs as an example, the corresponding expression in terms of residues reads
\begin{align}
&\left(\res{1,2,4,9}+\res{1,2,6,9}+\res{1,2,8,9}+\res{1,3,4,9}\right.\nnl
&\quad+\res{1,3,6,9}+\res{1,3,8,9}-\res{1,4,5,9}-\res{1,4,7,9}\nnl
&\left.\quad+\res{1,5,6,9}+ \res{1,5,8,9}-\res{1,6,7,9}+\res{1,7,8,9}\right)=0\,.
\end{align}
The above equality can be obtained by adding GRTs with the following source terms
\begin{equation}
 -(1,4,9)-(1,6,8)-(1,8,9)=0\,.
\end{equation}
Here and below the global minus sign in the equation will not be
noted for these vanishing results.

For finding a general rule describing which GRTs have to be added, it is
sufficient to restrict the attention to the case $i=1$, because all
other conformal constraints can be obtained by cyclical shifts. In the table below, 
a couple of lower-point examples are listed.
\begin{center}
\begin{small}
\begin{tabular}{cccl}
\hline
 part. & $\CE(i,k)$ & $m{=}k{-}i$ & source terms\\\hline
$7$     &   $\CE(1,3)$      &   2   &   $0=(7)$\\
$8$     &   $\CE(1,3)$      &   2   &   $0=(3,8)+(5,8)+(7,8)$\\
        &   $\CE(1,4)$      &   3   &   $0=(1,8)$\\
        &   $\CE(1,5)$      &   4   &   trivial vanishing\\
        &   $\CE(1,6)$      &   5   &   trivial vanishing\\
$9$     &   $\CE(1,3)$      &   2   &   $0=(3,4,9)+(3,6,9)+(3,8,9)+(5,6,9)+(5,8,9)+(7,8,9)$\\
        &   $\CE(1,4)$      &   3   &   $0=(1,4,9)+(1,6,9)+(1,8,9)$\\
        &   $\CE(1,5)$      &   4   &   $0=(1,2,9)$\\
        &   $\CE(1,6)$      &   5   &   trivial vanishing\\
        &   $\CE(1,7)$      &   6   &   trivial vanishing\\
$10$    &   $\CE(1,3)$      &   2   &   $0=(3,4,5,10)+(3,4,7,10)+(3,4,9,10)+(3,6,7,10)+(3,6,9,10)$\\
        &               &       &   $\,\,\,\,\,+(3,8,9,10)+(5,6,7,10)+(5,6,9,10)+(5,8,9,10)+(7,8,9,10)$\\
        &   $\CE(1,4)$      &   3   &   $0=(1,4,5,10)+(1,4,7,10)+(1,4,9,10)$\\
        &               &       &   $\,\,\,\,\,+(1,6,7,10)+(1,6,9,10)+(1,7,9,10)$\\
        &   $\CE(1,5)$      &   4   &   $0=(1,2,5,10)+(1,2,7,10)+(1,2,9,10)$\\
        &   $\CE(1,6)$      &   5   &   $0=(1,2,3,10)$\\
        &   $\CE(1,7)$      &   6   &   trivial vanishing\\
        &   $\CE(1,8)$      &   7   &   trivial vanishing\\
$\vdots$&   $\vdots$    &   $\vdots$ & $\quad\,\vdots$
\end{tabular}.
\end{small}
\end{center}
Based on those examples and further tests, the general rule for
obtaining the source terms in the fourth column for a certain $m$
and $n$ legs can be obtained as
\begin{equation}\label{eqn:confresult}
\sum_{\CV}(1,2,..,(m-2),\underbrace{v_{1},\ldots,v_{n-m-5}}_\CV,n)=0\,,
\end{equation}
where $\CV$ is a strictly increasing succession of $n-m-5$ numbers
\begin{equation}\label{eqn:vs}
v\in\{m+1,\ldots,n-1\},
\end{equation}
which has to be chosen such that the whole source term is strictly odd/even alternating, and the summation
is over all possible $\CV$s.

For example, in order to obtain the GRTs for the vanishing of
$\CE(1,5)$ in a scenario with $n=11$, one would start with
$(1,2,\CV, 11)$. According to~\eqn{eqn:vs} the numbers $v\in\CV$ have
to be in the range $\{5,\ldots,10\}$. Thus all valid choices for
$\CV$ in this scenario are
\begin{equation}
 (5,6),\,(5,8),\,(5,10),\,(7,8),\,(7,10),\,(9,10)\,,
\end{equation}
which finally leads to
\begin{align}
\CE(1,5)&=(1,2,5,6,11)+(1,2,5,8,11)+(1,2,5,10,11)\nnl&\,\,\,\,+(1,2,7,8,11)+(1,2,7,10,11)+(1,2,9,10,11)=0\,.
\end{align}

Considering the other type of dual conformal constraint (cf.
\eqn{eqn:conformaltree}),
\begin{equation}\label{eqn:conformaltree2}
\CE_{i,i-2}=-\CE_{i-1,i}=-2M^{\rm tree}_n\, ,
\end{equation}
one first needs to pick a form of the tree amplitude. For the investigation here it will be useful to choose
\begin{equation}\label{eqn:treeexpression}
 2M^\tree=M^\tree_{\rm BCFW}+M^\tree_{\rm P(BCFW)}\,,
\end{equation}
where the two representations of the tree-level amplitude have been
given in~\eqns{eqn:BCFW}{eqn:PBCFW}. The BCFW and the P(BCFW) form
of the tree amplitude are cyclically invariant, but in order to show
this, it is necessary to employ GRTs. For example, the equality
of the seven-point BCFW form of the tree amplitude to its shifted
version,
\begin{align}
&\res{2,3}+\res{2,5}+\res{2,7}+\res{4,5}+\res{4,7}+\res{6,7}\nnl
&=\res{3,4}+\res{3,6}+\res{3,1}+\res{5,6}+\res{5,1}+\res{7,1},
\end{align}
is not obvious. So we expect that only for one particular
choice of $i$ in~\eqn{eqn:conformaltree2} is the
expression~\eqn{eqn:treeexpression} obtained. Translating again
box coefficients into residues confirms this expectation. 
Only for $i=2$ is the chosen form of the tree amplitude
reproduced,
\begin{equation}\label{eqn:conftreeresult}
\CE(2,n)=-\CE(1,2)=-(M^\tree_{\rm BCFW}+M^\tree_{\rm P(BCFW)})\,.
\end{equation}

Having mapped all dual conformal constraints to sums of GRTs, it is
interesting to make contact to the IR equations. As discussed
in \cite{Brandhuber:2009xz}, any IR equation can
be written down as a particular combination of $\CE(i,k)$,
\begin{equation}\label{eqn:IRm=2}
\ki{i}_2:
\CE_{i,i+2}+\CE_{i+2,i}-\CE_{i+3,i}=-2M_n^{\rm
tree}\, ,
\end{equation}
and
\begin{equation}\label{eqn:IRm>2}
\ki{i}_m:
\CE_{i,i+m}+\CE_{i+m,i}-\CE_{i+1,i+m}-\CE_{i+m+1,i}=0\,,
\end{equation}
for $m=3,...,\lfloor n/2\rfloor$. It would be straightforward to
just add and subtract the appropriate source terms corresponding to
the different terms $\CE(i,k)$. However, because of cancellations
between different GRTs this does not result in the simplest
possible expression. Therefore the explicit analysis performed for
the dual conformal constraints will be repeated for the
IR equations below\footnote{Note that the considerations are again
limited to the starting point $i=1$, because one can always obtain results
for other $i$'s by cyclic shifts.}.

As an illustrations for IR equations, let's consider the kinematic
invariant $\ki{1}_2$ for the 7-point NMHV scenario. Either scanning
for the IR divergences or directly using \eqn{eqn:IRm=2} delivers
\begin{equation}
 C^{\rm 1m}_{1234}+C^{\rm 1m}_{7123}+\half C^{\rm 2mh}_{1235}-\half C^{\rm 2mh}_{6713}+\half C^{\rm 2mh}_{1236}
 -\half C^{\rm 2mh}_{3451}-C^{\rm 2me}_{7134}-\half C^{\rm 3m}_{3461}-\half
 C^{\rm 3m}_{7135}
\end{equation}
for the left hand side of~\eqn{eqn:IR}. Translating into residues, one obtains:
\begin{align}
(\res{7,1}+\res{5,1}+\res{3,1}+\res{4,5})
+(\res{6,7}+\res{4,7}+\res{2,7}+\res{3,4})\nnl
+\half(\res{2,5}+\res{5,6})
-\half(\res{7,3}+\res{3,4})
+\half(\res{2,3}+\res{3,6})
-\half(\res{4,5}+\res{5,1})\nnl
-(\res{7,1})-\half(\res{3,1})-\half(\res{7,5})
\end{align}
which by use of GRTs equals
\begin{equation}
\res{2,3}+\res{2,5}+\res{2,7}+\res{4,5}+\res{4,7}+\res{6,7}\,.
\end{equation}
This is precisely the BCFW-form of the 7-point NMHV tree-amplitude.

As a second example, let's consider the invariant $\ki{1}_5$ for 9
particles. The corresponding IR equation reads
\begin{align}
-\half C^{\rm 2mh}_{1235}-\half C^{\rm 2mh}_{5671}+ C^{\rm 2mh}_{9125}+ C^{\rm 2mh}_{4561}
-\half C^{\rm 2mh}_{8915}-\half C^{\rm 2mh}_{3451} -C^{\rm 2me}_{1245} -\half C^{\rm 3m}_{5681}+\half C^{\rm 3m}_{9135}\nnl
+\half C^{\rm 3m}_{4571} +C^{\rm 3m}_{1256} -C^{\rm 3m}_{5691}+ C^{\rm 3m}_{9145}
+\half C^{\rm 3m}_{1257}+\half C^{\rm 3m}_{4581}+\half C^{\rm 3m}_{1258} +\half C^{\rm 3m}_{5613}-\half C^{\rm 3m}_{9157}=0\,.
\end{align}
Translating into residues and sorting out the pre-factors (but not
using any GRT) leads to the following result:
\begin{align}
&\half\left(-\res{1,2,3,5}-\res{1,2,3,7}-\res{1,2,4,5}-\res{1,2,4,7}\right.\nnl
&\quad+\res{1,2,5,6}+\res{1,2,5,8}+\res{1,2,5,9}-\res{1,2,6,7}\nnl
&\quad+\res{1,2,7,8}+ \res{1,2,7,9}-\res{1,5,6,7}-\res{2,5,6,7}\nnl
&\quad\left.-\res{3,5,6,7}-\res{4,5,6,7}+\res{5,6,7,8}+\res{5,6,7,9}\right)=0\,.
\end{align}
It is not difficult to see that this equation arises by adding GRTs with the following source terms:
\begin{equation}
 -(1,2,5)-(1,2,7)-(5,6,7)=0\,.
\end{equation}

Now we turn to a general analysis of GRTs for IR equations. For $m=2$,
the linear combination of residues should coincide with the
expression for the tree amplitude. From
eqns. \eqref{eqn:confresult},\eqref{eqn:conftreeresult} and \eqref{eqn:IRm=2},
it is straightforward to see which GRTs lead to the
parity-invariant form of the tree amplitude
\eqn{eqn:treeexpression}. With other choices of GRTs, we can also
get BCFW, P(BCFW)(\eqns{eqn:BCFW}{eqn:PBCFW}) or any other
form of the tree amplitude in terms of residues.

In case of $m>2$ there is no infrared divergence on the right hand
side of \eqn{eqn:IR}. Therefore, the sum is expected to vanish by use
of certain combinations of GRTs.

The following table shows a couple of examples:
\begin{center}
\begin{small}
\begin{tabular}{ccl}
\hline
 particles & kin. inv & source terms\\\hline
$7$     &   $\ki{1}_3$  &   $0=(1)$\\
$8$     &   $\ki{1}_3$  &   $0=(1,4)+(1,6)$\\
        &   $\ki{1}_4$  &   $0=(1,2)+(5,6)$\\
$9$     &   $\ki{1}_3$  &   $0=(1,4,5)+(1,4,7)+(1,6,7)$\\
        &   $\ki{1}_4$  &   $0=(1,2,5)+(1,2,7)+(5,6,7)$\\
$10$    &   $\ki{1}_3$  &   $0=(1,4,5,6)+(1,4,5,8)+(1,4,7,8)+(1,6,7,8)$\\
        &   $\ki{1}_4$  &   $0=(1,2,5,6)+(1,2,5,8)+(1,2,7,8)+(5,6,7,8)$\\
        &   $\ki{1}_5$  &   $0=(1,2,3,6)+(1,2,3,8)+(1,6,7,8)+(3,6,7,8)$\\
$\vdots$&   $\vdots$    &   $\quad\,\vdots$\\
$12$    &   $\ki{1}_3$  &   $0=(1,4,5,6,7,8)+(1,4,5,6,7,10)+(1,4,5,6,9,10)$\\
        &               &   $\,\,\,\,\,+(1,4,5,8,9,10)+(1,4,7,8,9,10)+(1,6,7,8,9,10)$\\
        &   $\ki{1}_4$  &   $0=(1,2,5,6,7,8)+(1,2,5,6,7,10)+(1,2,5,6,9,10)$\\
        &               &   $\,\,\,\,\,+(1,2,5,8,9,10)+(1,2,7,8,9,10)+(5,6,7,8,9,10)$\\
        &   $\ki{1}_5$  &   $0=(1,2,3,6,7,8)+(1,2,3,6,7,10)+(1,2,3,6,9,10)$\\
        &               &   $\,\,\,\,\,+(1,2,3,8,9,10)+(1,6,7,8,9,10)+(3,6,7,8,9,10)$\\
        &   $\ki{1}_6$  &   $0=(1,2,3,4,7,8)+(1,2,3,4,7,10)+(1,2,3,4,9,10)$\\
        &               &   $\,\,\,\,\,+(1,2,7,8,9,10)+(1,4,7,8,9,10)+(3,4,7,8,9,10)$
\end{tabular}\label{tab:GRT}
\end{small}
\end{center}

Inspecting the above table, one initial observation can be made: any
one-loop IR equation for an $n$-point amplitude can be represented as
the sum of $n-6$ basic GRTs, each of which is a sum of $6$ residues.
However, some residues will cancel, thus the final number of terms is
smaller than $6\cdot(n-6)$.

Furthermore, based on the examples in the table above and
further tests up to $n=20$, we propose the general rule, which is
kept odd/even alternating, for $m=3$ IR equations,
\begin{equation}\label{eqn:genm2}
\ki{1}_3\,:\sum_{\CV}(1,\underbrace{v_{1},\ldots,v_{n-7}}_\CV)=0\,,
\end{equation}
where $\CV$ a strictly increasing succession of $n-7$ numbers,
\begin{equation}
\label{eqn:vs2} v\in\{4,\ldots,n-2\}.
\end{equation}



For $m>3$, another kind of source terms appears. While the type
already encountered for $m=3$ starts from the left with
$(1,\ldots)$, a second type has the form $(\ldots,n-2)$ starting
from the right. Again by testing examples up to $n=20$, for $4\leq
m\leq \lfloor n/2\rfloor$, we find the following general rule,
\begin{align}\label{eqn:generalform}
\ki{1}_m\,:\,
&\sum_{\CV}(1,2,\ldots,(m-2)\,,\underbrace{v_{1},\ldots,v_{n-m-4}}_\CV)\,\nnl
&\quad+\sum_{\CW}(\underbrace{w_{1},\ldots,w_{n-m-4}}_\CW\,,(m+1),\ldots,(n-2))=0\,,
\end{align}
where $\CV$ and $\CW$ are again strictly increasing successions of $n-m-4$ numbers satisfying
\begin{equation}\label{eqn:vsws}
m+1\leq v \leq n-2\and 1\leq w \leq m-2
\end{equation}
and chosen to respect the odd/even alternating structure of source terms. The
first term in~\eqn{eqn:generalform} is a generalization of~\eqn{eqn:genm2}.




%

The labels of source terms in our general formula~\eqn{eqn:generalform} are strictly
odd/even alternating, which nicely connects to the classification of residues: starting from
a strictly alternating source term, the residues in the resulting GRT will be either
strictly alternating themselves or have an ``all-but-one'' odd/even alternating structure such as $eoeoo$.
Translating this into the language of invariant labels and comparing with
the possible invariant labels singles out exactly types 1 to 4, as expected.


One of the most important implications
of one-loop IR equations, which is completely obscured in the BCFW
formalism is the remarkable identity. This identity relates the
BCFW representation of the tree-amplitude \eqn{eqn:BCFW} with the
P(BCFW) form \eqn{eqn:PBCFW},
\begin{equation}
 M_{\text{BCFW}}=M_{\text{P(BCFW)}}\,.
\end{equation}
Being highly nontrivial identities in the BCFW approach, they
have an astonishingly simple form in the language of residues.

While it was already shown in \cite{ArkaniHamed:2009dn} that
identities
\begin{equation}\label{eqn:remarkable}
 E\star O\star E\star\cdots=(-1)^{(n-5)} O\star E\star O\star\cdots
\end{equation}
are implied by GRTs, we have found a way to derive them
directly from combinations of GRTs.

The statement is simple: adding all GRTs corresponding to all source
terms of the form $oeoe...$ for a particular number of legs produces 
the remarkable identity
\begin{equation}\label{eqn:remind}
M_{\text{BCFW}}=M_{\text{P(BCFW)}}: O\star E\star O \cdots=0,
\end{equation}
where $O=\sum_{\rm k \, odd}(k)$ and $E=\sum_{\rm k \, even}(k)$,
and we have $n-6$ factors since this is a sum of GRTs labeled by
source terms. 

For example, adding GRTs with source-terms
\begin{equation*}
 (1,2),\,(1,4),\,(1,6),\,(1,8),\,(3,4),\,(3,6),\,(3,8),\,(5,6),\,(5,8)\,\,\text{and}\,\,(7,8)
\end{equation*}
produces the identity for $n=8$:
\begin{align}
&\res{2,3,4}+\res{2,3,6}+\res{2,3,8}+\res{2,5,6}+\res{2,5,8}\nnl
+&\res{2,7,8}+\res{4,5,6}+\res{4,5,8}+\res{4,7,8}+\res{6,7,8}\nnl
=-&(\res{1,2,3}+\res{1,2,5}+\res{1,2,7}+\res{1,4,5}+\res{1,4,7}\nnl
+&\res{1,6,7}+\res{3,4,5}+\res{3,4,7}+\res{3,6,7}+\res{5,6,7})
\end{align}

As expected by parity invariance, one obtains the same result by
adding GRTs corresponding to all source terms of the form
$eoeo...$\,.


\section{Conclusion}\label{sec:higherloop}

In this article we have identified which global residue theorems in
the Grassmannian formulation of \Nf SYM theory imply the recently
derived one-loop dual conformal constraints and the well-known one-loop
IR equations in the NMHV sector. For both sets of equations the
source terms for the
corresponding GRTs can be obtained from the general
rules~\eqns{eqn:confresult}{eqn:generalform}. In addition, the remarkable identity relating
the BCFW and the P(BCFW) form of the tree amplitude emerges from adding all GRTs with an
odd/even alternating pattern of source terms~\eqn{eqn:remind}.

According to the classification of NMHV residues performed in the initial 
part of the article, all one-loop residues are of odd/even alternating or ``all-but-one''
odd/even alternating structure. This nicely fits to the general rules: all GRTs
involved have strictly odd/even alternating source terms, which relate exactly these types of residues.

Even restricting the consideration to the NMHV sector, there are many
GRTs beyond the ones employed in the mapping. These are presumably
related to higher-loop dual conformal constraints and IR equations.
From the classification it is obvious which residues
contribute to higher-loop leading singularities. However, without a general 
formalism to single out an integral basis
for two loops and beyond, the identification of higher-loop dual
conformal constraints and IR equations is left for a future project.

Furthermore, although dual conformal constraints (and thus IR equations)
should be related to GRTs beyond the NMHV sector, a general map has
not been found so far. In~\cite{ArkaniHamed:2009dg}, general
contours for N${}^2$MHV tree amplitudes have been derived using ideas
from localization in the Grassmannian manifold. It would be extremely
interesting to generalize this analysis to leading singularities of
loop amplitudes beyond the NMHV sector. Once this is achieved, it should
be possible to identify the Grassmannian origin of dual conformal constraints
for $k>3$.

\section*{Acknowledgements} The authors would like to express their
gratitude to Nima Arkani-Hamed and Freddy Cachazo for numerous
conversations and for pointing our attention to
reference~\cite{Brandhuber:2009xz}. S.H. is grateful to Nima
Arkani-Hamed and IAS Princeton for hospitality during the final
stage of this work. The work of J.B. was supported by the
German-Israeli Project cooperation (DIP) and the German-Israeli
Foundation (GIF).


\end{document}